\documentclass{article}

\usepackage{amssymb,amsfonts,amsmath,stmaryrd}
\usepackage{cite,enumerate,float,indentfirst}
\usepackage{forloop}
\usepackage{stmaryrd}
\usepackage{color}
\usepackage{tikz}
\usetikzlibrary{arrows,snakes,backgrounds}
\usepackage{url}
\usepackage{hyperref}
\usepackage{textcomp}

\def\be{\begin{eqnarray}}
\def\ee{\end{eqnarray}}

\def\tr{{\rm tr}\,}
\def\Tr{{\rm Tr}\,}

\definecolor{red}{rgb}{1,0,0}
\definecolor{orange}{rgb}{1,0.5,0}
\definecolor{violet}{rgb}{0.7,0,1}

\hoffset= - 1.0in         

\newcount\hshift
\newcount\vshift
\newcount\tmp
\newcount\i
\newcount\j
\newcount\boxsize
\newcount\circleRadius
\newcount\xShift
\newcount\yShift
\newcount\len
\newcount\tmp
\newcount\prevElt
\newcount\curElt

\def\gridPut(#1,#2)#3{{
    \loccount\x
    \x=#1
    \multiply\x by \boxsize
    \loccount\y
    \y=#2
    \multiply\y by \boxsize
    \put(\x,\y){
      #3
    }
}}

\makeatletter
\newcommand\YD[1]{{
    \vshift=0
    \@for \elt:=#1 \do{
      \Yrow\vshift\elt
      \advance\vshift by -\boxsize
    }
}}
\makeatother

\def\Ybox{{
    \let\s\boxsize
    \put(\xShift, \yShift) {
      \put(0,0){\line(1,0){\s}}
      \put(\s,0){\line(0,1){\s}}
      \put(\s,\s){\line(-1,0){\s}}
      \put(0,\s){\line(0,-1){\s}}
    }
}}

\newcommand\Yrow[2]{{
    \hshift = 0
    \j = 0
    \loop \ifnum\j<#2
    \put(\hshift,#1){\Ybox}
    \advance\hshift by \boxsize
    \advance\j by 1
    \repeat
}}

\newcommand\Ycolumn[2]{{
    \vshift = 0
    \j = 0
    \loop \ifnum\j<#2
    \put(#1,\vshift){\Ybox}
    \advance\vshift by \boxsize
    \advance\j by 1
    \repeat
}}

\newcommand\youngEnv[1]{{
    \boxsize=10
    \circleRadius=5
    \xShift=0
    \yShift=0
    #1
}}

\makeatletter
\newcommand\inlineYD[1]{{
    \yShift=0
    \xShift=0
    \def\onlyFirst{\xShift=\elt \def\onlyFirst{}}
    \@for \elt:=#1 \do{
      \onlyFirst
      \advance\yShift by 1
    }
    \multiply\yShift by 5
    \multiply\xShift by 5
    \begin{picture}(\xShift,\yShift)(0,-\yShift)
      \youngEnv{\boxsize=5 \yShift=-5 \YD{#1}}
    \end{picture}
}}
\makeatother

\makeatletter
\newsavebox{\@brx}
\newcommand{\llangle}[1][]{\savebox{\@brx}{\(\m@th{#1\langle}\)}%
  \mathopen{\copy\@brx\kern-0.5\wd\@brx\usebox{\@brx}}}
\newcommand{\rrangle}[1][]{\savebox{\@brx}{\(\m@th{#1\rangle}\)}%
  \mathclose{\copy\@brx\kern-0.5\wd\@brx\usebox{\@brx}}}
\makeatother

\begin{document}

\title{\vspace{-1cm}{\Large {\bf
      The ``Null-A'' superintegrability for
      monomial matrix models
    }
    \date{}
}}

\maketitle
\vspace{-4.2cm}

\begin{center}
	\hfill ITEP/TH-13/22 \\
	\hfill IITP/TH-11/22 \\
	\hfill MIPT/TH-10/22
\end{center}

\vspace{0.7cm}

\begin{center}
\begin{large}S. Barseghyan$^{a}$\footnote{sergey.barseghyan06@gmail.com},
  A. Popolitov$^{a,b,c}$\footnote{popolit@gmail.com}
\end{large}
\end{center}

\begin{center}
  $^a$ {\small {\it Moscow Institute of Physics and Technology, Dolgoprudny 141701, Russia }} \\
  $^b$ {\small {\it Institute for Theoretical and Experimental Physics, Moscow 117218, Russia}}\\
  $^c$ {\small {\it Institute for Information Transmission Problems, Moscow 127994, Russia}}\\
\end{center}

\vspace{0.5cm}

\begin{abstract}
  We find that superintegrability (character expansion) property persists
  in the exotic sector of the monomial non-Gaussian matrix model, with potential
  $\Tr X^r$, in pure phase, where the naive partition function $\langle 1 \rangle$
  vanishes. The role of the (anomaly-corrected) partition function is played
  by $\left\langle\chi_\rho\right\rangle$ -- the Schur average of the suitably
  chosen \textit{square} partiton $\rho$; such partitions are well-known to correspond
  to singular vectors of the Virasoro algebra.
  Further, non-zero are only Schur averages $\left\langle \chi_\mu\right\rangle$
  for such $\mu$ that have $\rho$ as their $r$-core, and superintegrability formula
  features the value of the \textit{skew} Schur function $\chi_{\mu/\rho}$
  at special point. The associated topological recursion and Harer-Zagier formula
  generalizations so far remain obscure.
\end{abstract}

\bigskip

{\section{Introduction}\label{sec:introduction}
  Matrix models
  \cite{book:M-random-matrices,
    paper:IZ-the-planar-approximation-II,
    paper:MMM-generalized-matrix-models-as-conformal-field-theories-discrete-cases,
    paper:KMMMP-conformal-matrix-models-as-an-alternative,
    paper:M-integrability-and-matrix-models,
    paper:M-2d-gravity-and-matrix-models-I,
    paper:M-matrix-models-as-integrable-systems}
  can be thought of as quantum field theories in zero dimensions.
  As such, they can be used as a training ground, where complications of the full-fledged
  QFT already manifest themselves, yet, they occur in tame and more
  accessible form.

  Perhaps, the most interesting feature of any QFT is the presence of
  non-perturbative effects \cite{paper:M-string-theory-what-is-it}: behavior of the system in strong coupling regime,
  away from the Gaussian point, is not easily described in terms of small
  perturbations around the Gaussian point, even when one utilizes sophisticated
  resummation techniques. The proverbial example of a non-perturbative effect
  is the phenomenon of confinement and asymptotic freedom for quarks inside
  meson or baryon in quantum chromodynamics. This confinement/asymptotic freedom
  effective behavior is easily seen with numeric simulations
  \cite{paper:G-the-confinement-problem-in-lattice-gauge-theory},
  but direct theoretical explanation is so far missing.

  Understanding the essence of such effects in the simplified and controlled
  setting of matrix models is the main physical motivation for their study.
  Yet, direct approaches to non-perturbative effects in non-Gaussian matrix models
  are few. So far, the most notable progress has been made only for specially chosen
  Dotsenko-Fateev matrix models (multiple Selberg integrals)
  \cite{paper:MMS-on-the-df-representation-of-toric,
    paper:MMSh-conformal-blocks-as-dotsenko-fateev-integral-discriminants},
  that play crucial role in the AGT correspondence
  \cite{paper:MMS-proving-agt-as-hs-duality}.

  A natural wish is to broaden the class of non-Gaussian matrix models directly
  solvable away from the Gaussian point. On this route, the simplest class
  of models to start
  with are the monomial matrix models, which were studied in
  \cite{paper:CHPS-orbifolds-and-exact-solutions}.
  There it was discovered that direct solubility seems to persist in the
  non-Gaussian world in the form of \textit{superintegrability}
  (or, in other words, character expansion) -- the statement that
  (normalized) average of a character is again a character
  \begin{align}
    \left\langle character \right\rangle \sim character,
  \end{align}
  while more traditional integrable constructions: Ward identities
  and $\hat{W}$-operator (cut-and-join) representation require
  some fixes to become well-defined.
  
  Still, the consideration in \cite{paper:CHPS-orbifolds-and-exact-solutions}
  was restricted in two respects. First, the integration contours
  were chosen in a very specific and symmetric way (such choice of contours
  is called the \textit{pure phase}).
  Second, only ``physical'' subset of the possible matrix size $N$ was considered,
  namely, only such $N$ (in relation to the potential exponent $r$
  and contour number $a$) for which the partition function $Z = \left\langle1\right\rangle$
  is non-vanishing.

  Therefore, at least two possible generalization directions
  from \cite{paper:CHPS-orbifolds-and-exact-solutions} are possible.
  First, one can try to say something about arbitrary choice of contours,
  i.e. the monomial non-Gaussian model in \textit{mixed phase}.
  We do not pursue this line here and remain in the pure phase.
  Instead, we concentrate on the other generalization line.

  Namely, we study monomial matrix model in a pure phase,
  but $N$ and contour number $a$ are chosen in such a way that partition
  function is zero, but (non-normalized) average of some Schur functions
  is non-zero.

  For example we have (all necessary definitions can be found in Section~\ref{sec:trivia})
  \begin{align} \label{eq:first-example}
    r = 3,\ a = 1,\ N = r k + 2: & \\ \notag
    Z & \ \equiv \left\langle1\right\rangle =  0 \\ \notag
  \left\langle \chi_{[1]}
  \right\rangle & = \ 
  \frac{2}{3} \Gamma^2\left(\frac{1}{3}\right)
  \cdot
  \prod_{m=1}^l (-1) 3^3 \Gamma^3\left(m+1\right)
  \frac{\Gamma\left(m+1-\frac{2}{3}\right)
    \Gamma^2\left(m+1+\frac{1}{3}\right)
    \Gamma\left(m+1+\frac{2}{3}\right)
  }{\Gamma\left(m+1-\frac{1}{3}\right)}
  \end{align}

\noindent Such situations are simply absent for Gaussian theories and for
Gaussian (so-called Dijkgraaf-Vafa) phases of non-Gaussian matrix models.

We already see from example \eqref{eq:first-example} a promise
that the structure of non-zero answers is far from arbitrary
and indeed this is the case.
The most notable novel structural feature of the arising formulas
(see eqns. \eqref{eq:selection-as-special-schur} and \eqref{eq:schur-avg-exotic-explicit})
is the appearance of skew Schur function $\chi_{\mu/\rho}$, evaluated at special point,
on the right hand side.
\textbf{To the best of our knowledge, it is the first example where skew Schur functions enter the superintegrability formulas.}

The second notable feature is that only some (actually a small fraction of)
partitions $\mu$ have non-zero Schur average $\left\langle \chi_\mu\right\rangle$
for a given relation between $N$, $a$ and $r$. As it turns out these are
precisely such $\mu$ whose $r$-core $\rho$ is a \textbf{square partition}!
Since square partitions are well-known to correspond to singular vectors
for the Virasoro algebra (and its corresponding $(q,t)$-deformations)
this observation may pave the way to $(q,t)$-generalizations
of pure-phase monomial matrix models, which are so far missing
and proved to be difficult to obtain from simpler considerations.
At least, this is the possible complementary road to the search
of pure-phase $(q,t)$-generalizations via the tridiagonal
representation \cite{paper:MMP-from-superintegrability-to-tridiagonal-representation-of-beta-ensembles}.

Lastly, the non-normalized averages of the core square partitions $\rho$
are themselves fully factorized expressions (see \eqref{eq:core-average-explicit})
which suggests that the proof of our formulas along the lines of
(\cite{book:M-symmetric-functions-and-hall-polynomials}, examples 1,2,3 p.373)
is entirely possible.
Otherwise, superintegrability formulas \eqref{eq:core-average-explicit}
and \eqref{eq:schur-avg-exotic-explicit} can be straightforwardly obtained (proved)
via application of Theorem~6 from \cite{paper:CHPS-orbifolds-and-exact-solutions}
(see also remark after formula (4.49)).

From the physics' perspective the fact that, for certain choices of $N$,
the naive partition function $\left\langle1\right\rangle$ vanishes, but
the peculiar Schur average $\left\langle\chi_\rho\right\rangle \neq 0$ plays
partition function's role, can be thought of as a sort of \textit{anomaly}:
 for the theory to remain consistent and non-trivial for certain values
of coupling constants
\footnote{Where ``coupling constants'' are undertstood in broad sense,
also including the choice of integration contours.}
the ``bare'' and simple monomial action needs to receive
(seemingly ugly) logarithmic correction
\begin{align}
  \Tr X^r \longrightarrow \Tr X^r + \log \chi_\rho.
\end{align}
The resulting factorization for the normalized averages
$\left\langle\chi_\mu\right\rangle / \left\langle\chi_\rho\right\rangle$
is, from this perspective, a complete mystery. Therefore, this physics' perspective definitely
deserves to be better studied in the future.

\bigskip

On another, more philosophical, note for large potential exponent $r$
and specially chosen $N$ and $a$ the non-zero averages do not kick in
until the size of the Young diagram $\rho$ (and $\mu$) becomes $\gg r$
(see \eqref{eq:r-7-cores} and remark afterwards). This non-perturbative
non-Gaussian feature is reminiscent (and, as such, fascinating)
to the hypothetical strong-force non-linear effect
at the core of the fictional novel
``The world of Null-A'' \cite{book:V-the-world-of-null-a} plot.
This effect also didn't manifest itself until the very high degree of precision
(analog of the size of the Young diagram) was achieved.
It is also loosely parallel to the asymptotic freedom/confinement
phenomenon in QCD, where the strong effect is also triggered only starting from
certain distance.

\bigskip

The exact solvability, at least for the Gaussian matrix models
and Gaussian-like regimes of non-Gaussian matrix models
is strongly associated with constructions of topological recursion
and Harer-Zagier generating functions. The naive attempts to adapt
these constructions to the exotic sector of pure phase monomial matrix models
fail, as we briefly discuss in Section~\ref{sec:emergent-structures}.

\bigskip

We conclude (Section~\ref{sec:conclusion}) with possible future directions.




}

{\section{Trivia on monomial matrix models, their pure phases and exotic sector}
  \label{sec:trivia}

  Monomial matrix models are the simplest non-Gaussian Hermitian matrix models.
  They are integrals of expressions, involving Hermitian matrix $M$,
  that depend only on time-variables $p_k = \tr M^k = \sum_{i=1}^N \lambda_i^k$,
  where $\lambda_i$ are the eigenvalues of $M$. Namely, the integral (non-normalized average)
  of the function, say, $F\{p_k\}$ is given by
  \begin{align}
    \left\langle F\{p_k\}\right\rangle_{\vec\gamma} =
    \int_{\gamma_1}\dots\int_{\gamma_N} d\lambda_1\dots d\lambda_N
    \prod_{i<j}^N\left(\lambda_i-\lambda_j\right)^2
    \exp\left(-\sum_{i=1}^N\lambda_i^r \right) F\{p_k\},
  \end{align}
  where there is extra freedom (the novelty of non-Gaussian case) in selecting
  a set of integration contours $\gamma_1\dots\gamma_N$ so that the integral
  is convergent (see e.g. Section 3.2 of
  \cite{paper:CHPS-orbifolds-and-exact-solutions}
  for discussion).

  The average of a unit $Z := \left\langle 1 \right\rangle$ has special significance:
  motivated by thermodynamical analogy it is called \textit{partition function}
  and one usually considers normalized averages (correlators)
  \begin{align}
    \llangle F\{p_k\} \rrangle = \frac{1}{Z}\left\langle F\{p_k\}\right\rangle.
  \end{align}
  It is in terms of these normalized averages that various interesting emergent
  properties of matrix models are formulated: topological recursion
  \footnote{
  Normalization is essential only for ``algebro-geometric'' --
  Chekhov-Eynard-Orantin
  \cite{paper:CE-hermitian-matrix-model-free-energy,
    paper:CE-matrix-eigenvalue-model-feynman-graph-technique-for-all-genera,
    paper:CEO-free-energy-topological-expansion-for-the-2-matrix-model},
  Bouchard-Eynard
  \cite{paper:BE-think-globally-compute-locally}
  and Borot-Shadrin
  \cite{paper:BS-blobbed-topological-recursion}
  , formulations of topological recursion. The original Alexandrov-Mironov-Morozov
  \cite{paper:AMM-partition-functions-of-matrix-models-first-special-functions-of-string-theory,
    paper:AMM-solving-virasoro-constraints-in-matrix-models,
    paper:MM-check-operators-and-quantum-spectral-curves}
  (i.e. in terms of loop-insertion, or ``check'' operators)
  topological recursion and its recent Kontsevich-Soibelman
  \cite{paper:KS-airy-structures-and-symplectic-geometry-of-topological-recursion,
    paper:ABCO-the-abcd-of-topological-recursion}
  reformulation work for non-normalized quantities \textit{just as well}.
  }, Harer-Zagier representation and character expansion to name a few.

  The choice space of integration contours $\{\gamma_i\}$ is, in fact, finite-dimensional
  and corresponds to the Stokes sectors of the potential $V = \sum\lambda^r$.
  For generic choice of $\{\gamma_i\}$ answers for $\llangle F\{p_k\}\rrangle$
  are hopelessly complicated (i.e. the structure is unclear at the moment)
  and the most one can do is to build perturbation theory around minima
  of the potential (for this to work the potential $V$ itself must be perturbed
  to include only non-degenerate minima) -- this is the, so-called,
  Dijkgraaf-Vafa (DV) phase of the model
  \cite{paper:MSh-matrix-model-version-of-agt-conjecture-and-civ-dv-potential}.

  Such perturbative treatment makes HZ representation and character expansion
  (which are both finite $N$ properties) impossible, or nearly impossible,
  to discover and study (in novel situations).
  Therefore, on the quest to generalize these properties beyond Gaussian
  and Gaussian-like (doubly-logarithmic, long-tail
  \cite{paper:MMP-superintegrability-of-matrix-students-distibution}, etc.)
  cases one needs to develop a new approach.

  One such possible approach is to first choose $\{\gamma_i\}$ in a \textit{very}
  symmetric manner, in hope that finite $N$ answers become comprehendable
  and only then to gradually generalize from this verysymmetric (or, \textit{pure}) point.
  And indeed, if one chooses all $N$ contours to be equal to one and the same of $r$
  possible closed star-like integration contours
  \begin{align}
    C_{r,a} = \sum_{j=0}^{r-1} e^{-\frac{2 \pi j a}{r}} B_{r, j} \ \ , \ \
    B_{r, j} = \left[ 0, +\infty \cdot e^{2 \pi j/r} \right ) \ \ , \ \
      a = 0 \dots r-1,
  \end{align}
  both partition function $Z_a := Z_{\gamma_i = C_{r,a}}$ and
  (normalized) averages of Schur polynomials
  $\llangle \chi_\lambda \rrangle_a := \llangle \chi_\lambda \rrangle_{\gamma_i=C_{r,a}}$
  become amenable.

  We call such regime (all contours are identical and specially tuned)
  the \textit{pure phase} of the matrix model, or say that the model is in pure phase.
  Let us summarize, following \cite{paper:CHPS-orbifolds-and-exact-solutions},
  what is known about this regime (we ignore
  possible, but straightforward, $q$-generalizations for simplicity).

  The one-dimensional ($N=1$) non-normalized averages are equal
  \begin{align}
    \mathbb{M}_{r,a}(p) := \int_{C_{r,a}} d\lambda \exp\left(-\lambda^r\right) \lambda^p =
    \delta_{r|p+1-a} \Gamma\left(\frac{p+1}{r}\right),
  \end{align}
  where the divisibility Kronecker delta $\delta_{r|x}$ equals one when $x$ is
  divisible by $r$ and zero otherwise.

  Thanks to Jacobi-Trudi formula the non-normalized average of a Schur function
  (and partition function itself, as $1 \equiv \chi_\emptyset$)
  can be calculated as a determinant of a matrix made of these one-dimensional moments
  \begin{align}
    \left\langle\chi_\mu\right\rangle_a = N! \ 
    \det \phantom{}_{i,j=1}^N \ \mathbb{M}_{r,a}\left(\mu_{N-j+1}+i+j-2\right)
  \end{align}

  The partition function of Gaussian ($r=2$) MM,
  for its only non-trivial contour, has the form of product of factorials
  \begin{align}
    Z_{\text{Gaussian}} \sim \prod_{k=1}^N k!
  \end{align}

  The partition function of non-Gaussian ($r \neq 2$) MM
  and its non-normalized correlators, however, do \textbf{not}
  enjoy such simple dependence, even in pure phase.
  Instead, one has the splitting of the single answer
into $r$ branches, each corresponding to a particular value of the
remainder $b := \left\langle\frac{N}{r}\right\rangle$
(i.e. to obtain these answers one considers only $N$ of the form $k r + b$,
$k \in \mathbb{Z}$
and performs analytic continuation from just this subset).
\begin{align} \label{eq:pf-classic-pure}
  Z_{r,a} & = \frac{\delta_{r,a}(N)}{(2 \pi)^N}\prod_{i=0}^{N-1}
  \Gamma\left( \left\lfloor \frac{i}{r}\right\rfloor +1\right)
  \Gamma\left( \left\lfloor \frac{i-a}{r} \right\rfloor +\frac{a}{r} + 1\right )
  \\ \notag
  & \ \ \delta_{r,a}(N) = \left \{\begin{array}{ll}
  (-1)^{\left\lceil \frac{N}{r} \right\rceil \frac{a(a-1)}{2}
    + \left\lfloor \frac{N}{r} \right\rfloor \frac{(r-a)(r-a-1)}{2}},
  & \text{ if } N \equiv 0,a \text{ mod } r
  \\
  0 & \ \
  \text{otherwise}
  \end{array}
  \right . ,
\end{align}
so it turns out that partition function is non-zero only when
$\left\langle \frac{N}{r} \right\rangle = 0$ or $\left\langle \frac{N}{r} \right\rangle = a$. In this case, the normalized correlators can be defined
and the standard technique of orthogonal polynomials is applicable.
This is exactly the case that was studied in
\cite{paper:CHPS-orbifolds-and-exact-solutions}
and the resulting formula for the normalized Schur average is
\begin{align} \label{eq:schur-avg-classic-pure}
  \llangle \chi_\mu \rrangle_a = \frac{\delta_r(\mu)}{r^{|\mu|/r}}
  \prod_{(i,j)\in\mu}
  \frac{\llbracket N + j-i \rrbracket_{r,0}
    \llbracket N + j-i \rrbracket_{r,a}}
       {\llbracket h(i,j) \rrbracket_{r,0}},
\end{align}
where selection factor $\delta_r(\mu)$ is non-zero
and equal to the $r$-signature of the partition $\mu$ only when $\mu$
has trivial $r$-core (see, e.g. Appendix A in
\cite{paper:CHPS-orbifolds-and-exact-solutions}
for the definition of partition's core and signature)
and $h(i,j)$ is the hook length of the box $(i,j)$.
The special combinatorial quantity $\llbracket x \rrbracket_{r,s}$
selects certain diagonals in the Young diagram
from the product over all its boxes
\begin{align}
  \llbracket x \rrbracket_{r,s} =
  \left \{ \begin{array}{l} x, \text{ if } x \mod r = s \\
    1 \ \ \ \ \ \ \ \text{ otherwise }
  \end{array} \right . ,
\end{align}
therefore, non-trivially contributing diagonals of the Young diagram $\mu$
shift between the two cases $N = k r$ (i.e. $b=0$)
and $N = k r + a$ (i.e. $b = a$).

Already in \cite{paper:CHPS-orbifolds-and-exact-solutions} the observation was made that sometimes
(for particular choice of $r, a, b,$ and $\mu$) the non-normalized average
of a Schur function is non-zero even when the corresponding partition function
is zero. For instance, for $r = 3, a = 1, b = 2$ we have
\begin{align}
  \left\langle \chi_{[1]}
  \right\rangle = & \ 
  \frac{2}{3} \Gamma^2\left(\frac{1}{3}\right)
  \cdot
  \prod_{m=1}^l (-1) 3^3 \Gamma^3\left(m+1\right)
  \frac{\Gamma\left(m+1-\frac{2}{3}\right)
    \Gamma^2\left(m+1+\frac{1}{3}\right)
    \Gamma\left(m+1+\frac{2}{3}\right)
  }{\Gamma\left(m+1-\frac{1}{3}\right)}
\end{align}
but these cases were not studied exhaustively.
Yet, as $r$ grows, this \textit{exotic} sector $\left\langle \frac{N}{r} \right\rangle \neq 0, a$ becomes dominant: majority of $N$ have remainders modulo $r$ distinct from $0$ or $a$.

In this paper we do focus on this \textbf{exotic sector}.
It turns out that manifest expressions are not very different
from \eqref{eq:pf-classic-pure} and \eqref{eq:schur-avg-classic-pure}.
The suitable analog of the partition function is played by average
of certain ``base'' partition
(that depends on $r, a$ and $b$)
that are not unreasonable (and in fact very natural)
from the point of view of Virasoro algebra representation theory.

We summarize our results in the next section.


}

{\section{Answers in the exotic sector}
  \label{sec:exotic-answers}
  With help of \textbf{extensive} computer experiments we are able
  to establish the following.\footnote{These answers can be straightforwardly derived
  via application of Theorem~6 from \cite{paper:CHPS-orbifolds-and-exact-solutions},
  but computer experimentation is still faster.}

          Let us fix $a$ and $b$. Then the ``base'' partition, i.e.
          the first one for which the corresponding non-normalized
          Schur average is non-zero is the square partition
          (call it $\rho$) with sides
          \begin{align}
            \begin{array}{l}
              a > b: \ \ b \times (a - b) \\
              a < b: \ \ b-a \times (r - b) \\
            \end{array}
          \end{align}
          There is no rim hook of length $r$ that can be subtracted
          from this parititon, so it is equal to its $r$-core.

          The non-normalized average is manifestly equal to
          (we omit explicit specification of $r$, $a$ and $b$
          in the correlators from now on)
          \begin{align} \label{eq:core-average-explicit}
            \left\langle\chi_\rho\right\rangle =
            \left \{
            \begin{array}{l}
              a < b: \ \ \text{sign}(a,b,l)
              N!
              \Gamma\left(\frac{a}{r}\right)^a
              \Gamma\left(l{\color{red}+1} +\frac{a}{r}\right)^{(b-a)}
              \Gamma(l+1)^b
              \prod_{m=1}^l \Gamma\left(m + \frac{a}{r}\right)^r
              \Gamma\left(m\right)^r
              \\
              a > b: \ \ \text{sign}(a,b,l)
              N!
              \Gamma\left(\frac{a}{r}\right)^a
              \Gamma\left(l +\frac{a}{r}\right)^{(b-a)} \ \ \ \ 
              \Gamma(l+1)^b
              \prod_{m=1}^l \Gamma\left(m + \frac{a}{r}\right)^r
              \Gamma\left(m\right)^r
            \end{array}
            \right .
          \end{align}
          
          All the rest, ``descendent'' partitions $\mu$, which have
          non-zero Schur average are precisely the ones that
          have $\rho$ as their $r$-core.
          In this case the ratio of the averages is
          \begin{align}\label{eq:schur-avg-exotic-pure}
            \frac{\left\langle \chi_{\mu} \right\rangle}
                 {\left\langle \chi_{\rho} \right\rangle}
                 = \delta'_r(\mu) \frac{1}{r^{\frac{|\mu|-|\rho|}{r}}}
                 \prod_{(i,j)\in\mu}
                 \frac{\llbracket N + j-i \rrbracket_{r,0}
                   \llbracket N + j-i \rrbracket_{r,a}}
                      {\llbracket h(i,j) \rrbracket_{r,0}},
          \end{align}
          so the formula looks \textit{almost} the same as in the $Z \neq 0$ case
          \eqref{eq:schur-avg-classic-pure}.
          The care should be taken, however, to remember that since $N \equiv b \mod r$,
          the product is taken over the boxes with $i - j \equiv -b \mod r$
          and $i - j \equiv a-b \mod r$, respectively.
          It just remains to notice that
          selection/sign factors assemble with hook-length product to form
          Schur polynomial at special point
          \begin{align} \label{eq:selection-as-special-schur}
            \frac{\delta_r(\mu)}{\prod_{(i,j)\in\mu} \llbracket h(i,j) \rrbracket_{r,0}}
            = \chi_\mu\left(p_k=\delta_{k,r}\right);
            \ \ \ \ \
            \frac{\delta'_r(\mu)}{\prod_{(i,j)\in\mu} \llbracket h(i,j) \rrbracket_{r,0}}
            = \chi_{\mu/\rho} \left(p_k=\delta_{k,r}\right),
          \end{align}
          where $\chi_{\mu/\rho}$ is the skew-Schur polynomial.
          So, the formula for the (appropriately!) normalized Schur average
          in the exotic sector generalizes the formula in the usual sector
          in a very natural way.

          \bigskip

          The appearance of Schur polynomials, evaluated at special point,
          on the r.h.s. calls for an attempt to rewrite expressions
          \eqref{eq:schur-avg-classic-pure} and \eqref{eq:schur-avg-exotic-pure}
          \textit{completely} in terms of Schur functions.
          We do this in Section~\ref{sec:character-expansion}.
          
          \bigskip
          \bigskip
          \bigskip
          \bigskip
          

          Here is the table of all the base partitions, together
          with reminders, for $r = 7$
          \def\mybox#1#2#3 {
            \begin{array}{l}
              #1, #2 \\
              #3
            \end{array}
          }
          \begin{align} \label{eq:r-7-cores}
            \begin{array}{|c|c|c|c|c|c|c|}
              \hline
              a \slash b & 1 & 2 & 3 & 4 & 5 & 6 \\
              \hline
              0 & \mybox{-1}{6}{\inlineYD{6}} & \mybox{-2}{5}{\inlineYD{5,5}} &
              \color{red}\mybox{-3}{4}{\inlineYD{4,4,4}} & \mybox{-4}{3}{\inlineYD{3,3,3,3}} & \mybox{-5}{2}{\inlineYD{2,2,2,2,2}} & \mybox{-6}{1}{\inlineYD{1,1,1,1,1,1}}  \\
              \hline
              1 & \times & \mybox{-1}{5}{\inlineYD{5}} &
              \mybox{-2}{4}{\inlineYD{4,4}} &
              \mybox{-3}{3}{\inlineYD{3,3,3}} & \mybox{-4}{2}{\inlineYD{2,2,2,2}} &
              \mybox{-5}{1}{\inlineYD{1,1,1,1,1}} \\
              \hline
              2 & \mybox{-1}{1}{\inlineYD{1}} & \times & \mybox{-1}{4}{\inlineYD{4}} & \mybox{-2}{3}{\inlineYD{3,3}} &
              \mybox{-3}{2}{\inlineYD{2,2,2}} & \mybox{-4}{1}{\inlineYD{1,1,1,1}} \\
              \hline
              3 & \mybox{-1}{2}{\inlineYD{2}} & \mybox{-2}{1}{\inlineYD{1,1}} & \times &
              \mybox{-1}{3}{\inlineYD{3}} & \mybox{-2}{2}{\inlineYD{2,2}} &
              \mybox{-3}{1}{\inlineYD{1,1,1}} \\
              \hline
              4 & \mybox{-1}{3}{\inlineYD{3}} & \mybox{-2}{2}{\inlineYD{2,2}} &
              \mybox{-3}{1}{\inlineYD{1,1,1}} & \times & \mybox{-1}{2}{\inlineYD{2}} &
              \mybox{-2}{1}{\inlineYD{1,1}} \\
              \hline
              5 & \mybox{-1}{4}{\inlineYD{4}} & \mybox{-2}{3}{\inlineYD{3,3}} &
              \mybox{-3}{2}{\inlineYD{2,2,2}} & \mybox{-4}{1}{\inlineYD{1,1,1,1}} & \times
              & \mybox{-1}{1}{\inlineYD{1}} \\
              \hline
              6 & \mybox{-1}{5}{\inlineYD{5}} & \mybox{-2}{4}{\inlineYD{4,4}} &
              \mybox{-3}{3}{\inlineYD{3,3,3}} & \mybox{-4}{2}{\inlineYD{2,2,2,2}} &
              \mybox{-5}{1}{\inlineYD{1,1,1,1,1}}
              & \times \\
              \hline
            \end{array}
          \end{align}
          Notably, the growth of the maximal (over $a$ and $b$)
          number of boxes of the first non-zero partition
          is \textit{quadratic} in $r$. For instance,
          for $a = 0, \ b = 3$ this number is $12$, which means
          that averages of \textit{all} partitions
          with up to and including $11$ boxes vanish, and only
          at $12$ boxes some non-trivial contributions start.

          For $r = 10$, $a = 0$, $b = 4$ this becomes even more
          striking as first non-vanishing contribution is
          on the level $28 >> 2 r$ (!). Such is the nature
          of the truly non-perturbative strongly coupled physics.
}
{\section{Character expansion}\label{sec:character-expansion}
  
  Certain matrix models possess a very peculiar property -- \textit{superintegrability}.
  This property means that values of certain normalized correlators are particularly simple
  and, very informally, can be stated as
  \begin{align}
    \llangle character \rrangle \sim character,
  \end{align}
  that is, the normalized average of a character (in cases we consider -- Schur function)
  is equal to the (product of) characters, evaluated at certain points.
  
  At the moment the story is in active development, so it is not clear how to tell
  in advance, whether a given matrix integral is superintegrable,
  or what is the meaning behing superintegrability: it is discovered
  case by case (see \cite{paper:MM-superintegrability-summary} for recent review).
  Therefore, every advance in this direction is valuable.
  

  The appearance of characters (Schur polynomials) evaluated at special point
  \eqref{eq:selection-as-special-schur} in formulas for normalized Schur average
  \eqref{eq:schur-avg-classic-pure} and \eqref{eq:schur-avg-exotic-pure}
  is very telling: a natural next question is whether it is possible to rewrite the r.h.s
  of \eqref{eq:schur-avg-classic-pure} and \eqref{eq:schur-avg-exotic-pure}
  entirely as a product of Schur functions, each evaluated at a certain point.

  Indeed, for the Gaussian ($r=2$) case the relevant
  rewrite is straightforward
  \begin{align} \label{eq:gaussian-character-expansion}
    \llangle \chi_\mu \rrangle
    = \chi_\mu (p_k=\delta_{k,2}) \frac{\chi_\mu(p_k = N)}{\chi_\mu (p_k = \delta_{k,1})}
  \end{align}
  for its only non-trivial choice of integration contour.

  In fact, the generalization of the superintegrability property
  to the $r \neq 2$ case is already present in \cite{paper:CHPS-orbifolds-and-exact-solutions}, theorem 4.
  However, firstly, the proof is only done for the regular case
  $N \equiv 0,a \text{ mod } r$
  and not the exotic case, secondly, average of the Schur polynomial
  in the orbifold model is just rewritten through the Schur averages
  in the simpler ($r = 1$) models and not through Schur polynomials at special points,
  though it is straightforward to do so.
  Lastly, the formula (4.14) of \cite{paper:CHPS-orbifolds-and-exact-solutions} features peculiar shifts
  of Fayet-Illiopoulos (i.e. insertion of the deteminant) terms.

  Here we simplify all this into a much more explicit formula. Namely,
  \begin{align}\label{eq:schur-avg-exotic-explicit}
    \frac{\left\langle \chi_\mu \right\rangle}
         {\left\langle \chi_\rho \right\rangle}
         = & \ \frac{1}{r^{(|\mu| - |\rho|)/r}
           \chi_{\mu/\rho}(p_k=\delta_{k,r})} \\ \notag
         & \times
         \prod_{j=0}^{r-1}
         \chi_{\mu^{\left(m_r(j + m_r(b-a))\right)}}
         \left(p_k = \frac{1}{r}\left( N
         + \delta_{j < b}\cdot m_r(-b) + \delta_{j \geq b}\cdot n_r(-b) \right)
         \right )
         \\ \notag & \times
         \prod_{j=0}^{r-1}
           \chi_{\mu^{(m_r(j+b))}}
           \left(p_k = \frac{1}{r}\left( N +
           \delta_{j < m_r(b-a)}\cdot n_r(a-b) + \delta_{j \geq m_r(b-a)}\cdot m_r(a-b)
           \right)\right)
  \end{align}
  where the $\mu^{(i)}$, $i = 0 \dots r-1$ are the $r$-quotients
  of the diagram $\mu$ (that can be, for instance, be read from the corresponding
  \textit{abacus diagram}, see Appendix A in \cite{paper:CHPS-orbifolds-and-exact-solutions}
  and \cite{book:M-symmetric-functions-and-hall-polynomials}[pp.12-14]).
  The functions $m_r$ and $n_r$ are operations of taking the remainder modulo $r$
  in the specified range, even for the negative value of the argument, that is
  \begin{align}
    m_r(x) = & \ x \mod r \in [0, r-1] \text{ and } \\ \notag
    n_r(x) = & \ m_r(x) - r \in [-r, -1]
  \end{align}

  Note that, firstly, the skew Schur function $\chi_{\mu/\rho}$
  is now in the denominator (c.f. \eqref{eq:schur-avg-exotic-pure},
  \eqref{eq:selection-as-special-schur}).
  This is because it is the selective product over boxes
  $\prod \llbracket N + \dots \rrbracket$ times the special skew Schur
  $\chi_{\mu/\rho}$ that is equal to the product over
  Schurs of $r$-quotients.
  
  Secondly, the functions $m_r(x)$ and $n_r(x)$ are necessary to
  accomodate all of the $2 \times 2 = 4$ possible regions in the remainder
  space.

  Thirdly, it would be highly desirable to find such a form of
  \eqref{eq:schur-avg-exotic-explicit} that does not include
  the $r$-quotients at all -- only, perhaps, skew Schur functions
  at special points. At the moment, however, we are unaware of such
  expression and its search is one of the topics for future research.
}


{\section{(Failure to find) Emergent structures}\label{sec:emergent-structures}
  Once we have the superintegrability formulas, we can in principle
  obtain any (non-normalized) multitrace correlator by summing Schur
  averages with appropriate coefficients. Then we can assess whether
  some structures that are formulated in terms of these multitrace
  correlators and are usually associated with integrability and solvability
  persist in the exotic sector.

  However, we immediately face the problem. For most cases
  (when main diagonal of the core partition $\rho$ has two or more boxes)
  the single trace correlators all vanish, since they are made only from
  single hook Schur functions
  \begin{align}
    p_k = \sum_{d=0}^{k-1} (-1)^d \chi_{[k-d, 1^d]}(p)
  \end{align}
  
  \bigskip
  
  Therefore the definition of multitrace correlators should be
  \textit{modified} for the exotic sector if one hopes to get non-trivial
  answers.

  As the first naive attempt (which we soon show to be a failed one)
  we can define the \textit{exotic analogue} of a single-trace correlator
  to be
  \begin{align}\label{eq:modified-singletrace}
    \llangle \Tr M^k \rrangle_{exotic} :=
    \sum_{d=0}^{k-1} (-1)^d
    \frac{\chi_{F_1(\rho,\mu=[k-d,1^d], r)}}{\chi_\rho},
  \end{align}
  where $F_1(\rho,\mu,r)$ is such unique partition that has $\rho$
  as its $r$-core and $\mu$ as its first $r$-quotient (and the rest
  of its $r$-quotients are trivial partitions). For instance we have
  \begin{align}
    F_1([2],[2,1], 3) = [3, 3, 1, 1, 1, 1, 1]
  \end{align}

  With this modified definition this is what becomes of two traditional
  integrable structures.
  
  {\subsection{Topological recursion} \label{sec:top-rec}
    Let us consider the case of (modified \textit{a la} \eqref{eq:modified-singletrace})
    single-trace correlator for $r=7$, $a=0$ and $b=2$.
    Then the non-trivial core partition $\rho = [5,5]$ and we get
    \begin{align} \label{eq:tr-failing-correlators}
      \llangle \Tr X^1 \rrangle_{exotic} = & \ 
      -\frac{4}{49} + \frac{4 }{49} N - \frac{1}{49}N^2
      \\ \notag
      \llangle \Tr X^2 \rrangle_{exotic} = & \ 
      \frac{16}{343} - \frac{24}{343} N + \frac{12}{343} N^2 - \frac{2}{343} N^3
      \\ \notag
      \llangle \Tr X^3 \rrangle_{exotic} = & \ 
      -\frac{276}{2401} + \frac{356}{2401} N - \frac{169}{2401} N^2
      + \frac{40}{2401} N^3 - \frac{5}{2401} N^4
    \end{align}

    We see that the basic initial assumption of the topological recursion
    -- that the powers of $N$ decrease by 2, is not satisfied here.
    Hence the correlators \eqref{eq:tr-failing-correlators} do not satisfy
    topological recursion in the narrow (Chekhov-Eynard-Orantin) sense
    \cite{paper:CE-hermitian-matrix-model-free-energy,
      paper:CE-matrix-eigenvalue-model-feynman-graph-technique-for-all-genera,
      paper:CEO-free-energy-topological-expansion-for-the-2-matrix-model}.
    Whether they satisfy more flexible versions, like ``blobbed''
    topological recursion of Borot-Shadrin \cite{paper:BS-blobbed-topological-recursion}
    or the universal check-operator formalism of Alexandrov-Mironov-Morozov
    \cite{paper:AMM-partition-functions-of-matrix-models-first-special-functions-of-string-theory,
      paper:AMM-solving-virasoro-constraints-in-matrix-models,
      paper:MM-check-operators-and-quantum-spectral-curves},
    recently rebranded as Kontsevich-Soibelman topological recursion
    \cite{paper:KS-airy-structures-and-symplectic-geometry-of-topological-recursion,
      paper:ABCO-the-abcd-of-topological-recursion}
    is a question for separate extensive research.

    Moreover, by looking at first correlators for same $r$ and different $a$ and $b$
    \begin{align}
      \scriptstyle
      r = 3, a = 0, b = 1: \ \rho = [2] & \scriptstyle \ \
      \llangle \Tr X \rrangle_{exotic} =
      \frac{1}{9}N^2-\frac{2 }{9}N+\frac{1}{9}
      , \ \
      \llangle \Tr X^2 \rrangle_{exotic} =
      -\frac{1}{81}N^4+\frac{4 }{81}N^3-\frac{5}{27}N^2
      +\frac{22 }{81}N-\frac{10}{81}
      \\ \notag
      \scriptstyle
      r = 3, a = 0, b = 2: \ \rho = [1,1] & \scriptstyle \ \
      \llangle \Tr X \rrangle_{exotic} =
      -\frac{1}{9}N^2-\frac{2 }{9}N-\frac{1}{9}
      , \ \
      \llangle \Tr X^2 \rrangle_{exotic} =
      -\frac{1}{81}N^4-\frac{4 }{81}N^3-\frac{5}{27}N^2-\frac{22 }{81}N-\frac{10}{81}
      \\ \notag
      \scriptstyle
      r = 3, a = 1, b = 2: \ \rho = [1] & \scriptstyle \ \
      \llangle \Tr X \rrangle_{exotic} =
      \frac{1}{9}N^2-\frac{1}{3}N+\frac{2}{9}
      , \ \
      \llangle \Tr X^2 \rrangle_{exotic} =
      -\frac{1}{81}N^4+\frac{2 }{27}N^3-\frac{22}{81}N^2+\frac{13}{27}N-\frac{22}{81} 
    \end{align}
    we see that genus zero contribution is actually the same (modulo sign).
    It could be that properly adjusted for the exotic sector analogs of the
    genus zero two-point functions are actually different in these cases,
    and this is what allows to reproduce from topological recursion the difference
    between higher genus corrections,
    however, we also postpone the detailed analysis for the future research.
  }

  {\subsection{Harer-Zagier generating functions} \label{sec:harer-zagier}
    The manifestation that Harer-Zagier generating functions
    \cite{paper:HZ-the-euler-characteristic-of-the-moduli-space-of-curves,
      paper:MSh-from-brezin-hikami-to-harer-zagier-formulas-for-gaussian-correlators}
    are present for a given matrix model is that the Laplace transform
    of the correlators nicely factorizes
    \cite{paper:MPSh-quantization-of-harer-zagier-formulas,
      paper:MMPSh-harer-zagier-formulas-for-knot-matrix-models}.
    Taking \eqref{eq:modified-singletrace} as the definition of the single-trace
    correlators we immediately see that in a number of simplest examples
    there is no factorization of putative HZ-correlators
    \begin{align}
      r = 3, \ a = 1, \ b = 2: & \ \ \rho = [1] \\ \notag
      \llangle \Tr X \rrangle_{\text{HZ}} & \ := \sum_{N=0}^\infty \lambda^N
      \llangle \Tr X \rrangle_{exotic}(N) =
      -\frac{2 \left(3 \lambda^2-3 \lambda+1\right)}{9 (\lambda-1)^3}
      \\ \notag
      \llangle \Tr X^2 \rrangle_{\text{HZ}} & \ =
      \frac{2 \left(45 \lambda^4-99 \lambda^3+110 \lambda^2-55     
        \lambda+11\right)}{81 (\lambda-1)^5}
      \\ \notag
      \llangle \Tr X^3 \rrangle_{\text{HZ}} & \ =
      -\frac{40 \left(162 \lambda^6-567 \lambda^5+1028 \lambda^4-1040
        \lambda^3+609 \lambda^2-203 \lambda+29\right)}{2187 (\lambda-1)^7}
    \end{align}

    \bigskip
    
    Hence, some further thought is required to establish how this classic construction
    generalizes to exotic sector of monomial matrix models.
  }

}

{\section{Conclusion}\label{sec:conclusion}
  In this paper we provide more evidence (in the form of manifest expressions
  \eqref{eq:schur-avg-exotic-explicit} and \eqref{eq:core-average-explicit})
  that superintegrability is indeed the right language to express
  exact solvability in fully non-perturbative, essentially non-Gaussian
  regime of matrix models.
  We conclude with several promising future directions:
  \begin{itemize}
  \item $\beta$-deformation for monomial matrix models (in the form of an integral)
    that gives nicely factorized superintegrable expressions
    is not known at the moment.
    The difficulty seems to be in selection of the correct branch
    of the (fractional) power of the van der Monde determinant.
    Since Jack polynomials of square Young diagrams are also singular
    vectors of Virasoro algebra for $\beta \neq 1$, one can postulate
    the straightforward Jack generalization for formula \eqref{eq:schur-avg-exotic-explicit},
    restore the corresponding Ward identities and, hopefully,
    recover the correct integral definition, with proper branch selections.
    This, then, would imply the correct form of the $\hat{W}$-operator,
    and the proof of thus obtained superintegrability formulas
    would be possible along the lines of
    \cite{paper:MO-superintegrability-in-beta-deformed-gaussian-hermitian-matrix-model-from-w-operators}.
    
  \item The ``physical'' and the ``exotic'' sectors of monomial matrix models
    in pure phase together allow to write down integral representation
    for $(N - i + j)$-product over selected diagonals, where separation
    between first and second non-trivial diagonals is $n_1$, between second and third
    $n_2$, between third and fourth $n_1$ again and so on, so non-negative integer
    gaps $n_1$ and $n_2$ alternate.
    Let us call such a product \textit{double diagonal product}.
    From purely combinatorial viewpoint one may wonder, whether it is possible
    to write an integral representation for \textit{single} diagonal product
    (gaps are $(\dots,n_1,n_1,n_1,\dots)$), \textit{triple} diagonal product
    (gaps are $(\dots,n_1,n_2,n_3,n_1,n_2,n_3\dots)$), quadruple and so on.
    Thus obtained families of matrix models may well be the candidates for
    universal building blocks of non-Gaussian answers.
  \item Having understood the pure phase of monomial matrix model in full generality
    (i.e. both physical and exotic sector) one may pose the question
    of relevant universal $\hat{W}$-operator representation. It would be particularly
    interesting to see how different branches (both in contour number $a$ and
    $\left\langle\frac{N}{r}\right\rangle$ remainder $b$) are distinguished
    in this $\hat{W}$-operator form.
    
  \item Recently it was discovered that superintegrability is present
    well beyond ordinary matrix models. In the form of averages of Q-Schur functions it is present for the celebrated Kontsevich model
    \cite{paper:MM-superintegrability-of-kontsevich-matrix-model,
      paper:MM-superintegrability-and-kontsevich-hermitian-reltion}
    and it also generalizes to fermionic analogs of matrix integrals
    \cite{paper:MMMZh-natanzon-orlov-model-and-refined-superintegrability}.
    It would be very interesting to see, whether non-gaussian loci of
    these modes, or perhaps their suitable generalizations, also are
    superintegrable.
    
  \item The quesions of proper topological recursion and Harer-Zagier
    definitions are also present, yet, it appears that additional physical ideas
    are required to arrive at the correct versions.
  \end{itemize}
}

\section*{Acknowledgements}

This work was supported by the Russian Science Foundation (Grant No.20-12-00195).

\bibliographystyle{mpg}
\bibliography{references_shadow-phase-mon-matr}

\end{document}